\documentclass[reprint,showpacs,preprintnumbers,amsmath,amssymb, aip]{revtex4-1}
\usepackage{graphicx,wasysym}
\usepackage{dcolumn}
\usepackage{bm}
\usepackage{tabulary}
\newcommand{\kB}{k_{\mathrm{B}}}
\usepackage{natmove}
\begin{document}

\preprint{1}

\title{Water exchange at a hydrated platinum electrode is rare and collective}

\author{David T. Limmer}
\affiliation{%
Princeton Center for Theoretical Science, Princeton University, Princeton, NJ, USA 08540
}%
\email{dlimmer@princeton.edu}
\author{Adam P. Willard}
\affiliation{%
Department of Chemistry, Massachusetts Institute of Technology, Cambridge, MA, USA 94609
}%
\email{awillard@mit.edu}
\author{Paul A. Madden}
\affiliation{%
Department of Material Science, Oxford University, Oxford, UK
}%

\author{David Chandler}
\affiliation{%
Department of Chemistry, University of California, Berkeley, CA, USA 94609
}%

\date{\today}
\begin{abstract}
We use molecular dynamics simulations to study the exchange kinetics of water molecules at a model metal electrode surface -- exchange between water molecules in the bulk liquid and water molecules bound to the metal.  This process is a rare event, with a mean residence time of a bound water of about 40 ns for the model we consider.  With analysis borrowed from the techniques of rare-event sampling, we show how this exchange or desorption is controlled by (1) reorganization of the hydrogen bond network within the adlayer of bound water molecules, and by (2) interfacial density fluctuations of the bulk liquid adjacent to the adlayer. We define collective coordinates that describe the desorption mechanism. Spatial and temporal correlations associated with a single event extend over nanometers and tens of picoseconds. 
\end{abstract}

\maketitle

\section{Introduction}
Hydrated metal interfaces are standard platforms for heterogenious catalysis~\cite{Somorjai:2010p8235}.  Understanding their microscopic kinetic processes is important for the design of efficient and clean catalytic systems~\cite{Nrskov:2009p7956}. Currently, this understanding is limited by too little knowledge of the molecular-level structure and dynamics at interfaces. Most experimental work has come from macroscopic electrochemical measurements where inferring molecular-level details is challenging~\cite{bard2001electrochemical,vetter1967electrochemical}, or surface science measurements at ultrahigh vacuum conditions, where processes relevant to condensed phase behavior are absent~\cite{verdaguer2006molecular,hodgson2009water}. Most theoretical work has utilized quantum chemical methods designed to produce highly accurate representations of one (or a few) configurations, like that reviewed in Refs.~\onlinecite{Nrskov:2009p7956} and \onlinecite{Carrasco:2012p3354}.   That type of study neglects fluctuations, relaxation and solvation.  Here, we adopt a complimentary perspective, following in the footsteps of Refs.~\onlinecite{straus1995calculation,spohr2002molecular,wilhelm2010proton,rose1991solvation,mattsson2003methanol,raghavan1991structure,rose1993adsorption}, employing classical molecular dynamics simulations to study large-length scale correlations and long-time scale dynamics.  

We use these methods to study the dynamics of exchange between water molecules in the bulk liquid with water molecules tightly bound to a platinum electrode.  We show how the bound water molecules organize into an adlayer with long-lived disordered spatial patterns extending over large distances. Similarly, we show how the bulk water molecules adjacent to the monolayer form a soft liquid interface that can accommodate long wavelength distortions in the local interfacial density.  The exchange between the bulk and the bound waters involves the dynamics of both the adlayer and the soft liquid interface.  The exchange process is rare, occurring with a characteristic timescale on the order of tens of nanoseconds.  It is collective, proceeding through a reaction coordinate that involves reorganization of many adlayer molecules in concert with fluctuations of the soft liquid interface. 

This work highlights that in the vicinity of an aqueous electrode interface, even the simplest kinetic processes become highly collective.  It is in line with our previous work that showed how an aqueous electrode interface exhibits an array of emergent behaviors~\cite{willard2013characterizing,limmer2013hydration}. In particular, large binding energies pin water molecules to the surface resulting in a water adlayer in which hydrogen bonding is passivated almost entirely within the plane of the electrode. This in-plane hydrogen bonding greatly reduces interactions between the adlayer and surrounding liquid, producing a composite water-metal surface that is hydrophobic.  The composite surface attracts defects from the bulk liquid like excess protons,\cite{limmer2013hydration,cao2015hydrated} and weakly solvated ions\cite{limmer2015nanoscale}, and it facilitates large density fluctuations in the adjacent bulk liquid~\cite{limmer2013hydration}. Further, when surface periodicities and local coordinations of the electrode surface are incommensurate with favorable hydrogen bonding geometries, hydrogen-bonding patterns of the adlayer are frustrated, leading to structures with localized defects and heterogeneous dynamics~\cite{willard2013characterizing}.  

Consequently, the dynamics at this interface is governed by a hierarchy of timescales, spanning picoseconds to tens of nanoseconds.  The shortest of these timescales characterize molecular fluctuations in the bulk liquid, the longest characterize relaxation of the adsorbed water monolayer~\cite{willard2013characterizing}. These fast and slow degrees of freedom are coupled through fluctuations of the bulk-monolayer interface that itself relaxes over intermediate timescales~\cite{limmer2013hydration}.  This complexity affects the exchange of a water molecule between the adsorbed monolayer and the bulk, and as such, it also affects the accessibility of reactive sites on a hydrated electrode. Recognizing this fact, Hansen et al introduce phenomenological energy barriers in their treatment of oxygen evolution~\cite{hansen2014unifying}. These barriers are attributed to entropic effects related to surface water reorganization, highlighting the role of slow interfacial water dynamics. Related effects may be important in gas phase catalysis as well because a similar monolayer of water forms on platinum surfaces exposed to ambient air.  Slow reorganization of this monolayer has been conjectured to account for the low sticking coefficient of gaseous adsorbates such as carbon dioxide~\cite{lofgren1998poisoning,standop2012h2o}, decreasing catalytic efficiency. 

\begin{figure}[t]
\begin{center}
\includegraphics[width=8.5cm]{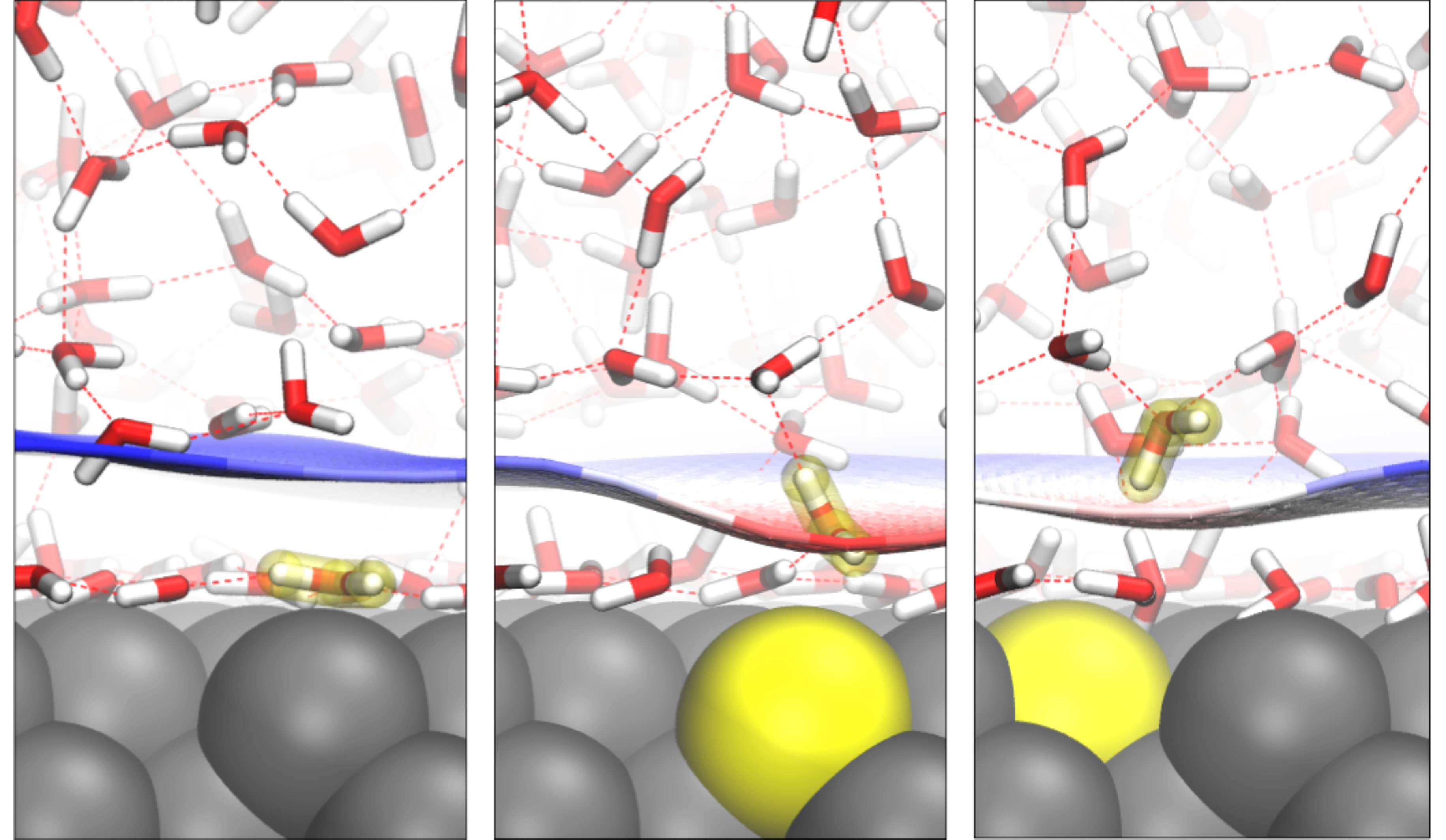}
\caption{(left) Characteristic snapshot the adsorbed state. The tagged molecule, highlighted in yellow, sits atop a platinum atom (grey), below the instantaneous interface (blue), and hydrogen bonds to other waters (red and white) predominately within the adlayer plane. (center) Characteristic snapshot of a member of the transition state ensemble. A rare density fluctuation in the liquid above, rendered with the deformed instantaneous interface, allows for the tagged molecule to come off of the electrode, leaving a vacancy on the surface, shown in yellow. (right) Characteristic snapshot of the desorbed state. The tagged molecule is above the instantaneous interface and the vacancy diffuses away. Snapshots are separated by 1 ps. A movie of the same trajectory can be streamed online at \small{https://youtu.be/MVjUBTBrmfU}.}
\label{Fi:1}
\end{center} 
\end{figure}

Figure \ref{Fi:1} illustrates the basic steps in the first half of a typical exchange process.  This first half is the desorption of a water molecule from the electrode surface as found in the simulations we detail in the next sections of this paper. The three panels of the figure are each separated by approximately 1 ps, with a tagged molecule initially adsorbed to the metal, proceeding through its transition state where it leaves behind a surface vacancy, before committing to bulk liquid where the vacancy either diffuses away or becomes annihilated by an adsorption event. The orientational relaxation of water molecules within the adlayer occurs roughly $10^4$ times more slowly then for those within the bulk liquid, and nearly $10^3$ times more slowly than at the bulk liquid interface. Not surprisingly, a detailed characterization of molecular desorption requires the quantification of local surface reorganization ability. 

Figure \ref{Fi:1} also renders the liquid's instantaneous interface adjacent to the composite electrode-water surface. This soft-liquid interface is identified as in Ref.~\onlinecite{Willard:2010p1954}. We see below that it is important in distinguishing the adsorbed and desorbed states with reference to the position of the instantaneous liquid phase boundary.  Only with this frame of reference does it become clear that desorption is an instantonic event with few recrossings.  Otherwise the nature of the process is obscured by slow capillary-wave-like dynamics of the interface.  With the adsorbed and desorbed states or basins for a tagged water molecule thus defined, we use tools of transition path sampling~\cite{bolhuis2002transition} to consider the mechanism of desorption.  In particular, we determine a collective reaction coordinate that describes the ensemble of transition states. Finally, we consider the dynamics involved in the time-reverse process, i.e., the adsorption of a molecule from the bulk onto the electrode. The Appendix provides additional results and details on our methodology.

\section{Basin definitions and timescales}\label{Sec:Time}

In order for an adsorbed molecule to enter the bulk liquid it must incorporate itself into the established hydrogen bonding environment of the liquid. Such a process is initiated through the scavenging of broken hydrogen bonds that arise spontaneously but fleetingly at a liquid water interface. This process is thus modulated by spatial fluctuations in the position of the bulk liquid interface that, due to the hydrophobic nature of the extended water-metal interface\cite{limmer2013hydration}, are large and driven by collective displacements of weakly-bound molecules. Consequently if the position of the desorbing molecule is described relative to a static frame of reference, the fluctuations in the soft liquid interface are likely to obscure meaningful details related to the formation of the initial hydrogen bonds between a desorbing molecule and the bulk. In order to account for these fluctuations explicitly, we adopt a reduced description of the instantaneous density field that incorporates these fluctuations.

\begin{figure}
\begin{center}
 \includegraphics[width=8.5cm]{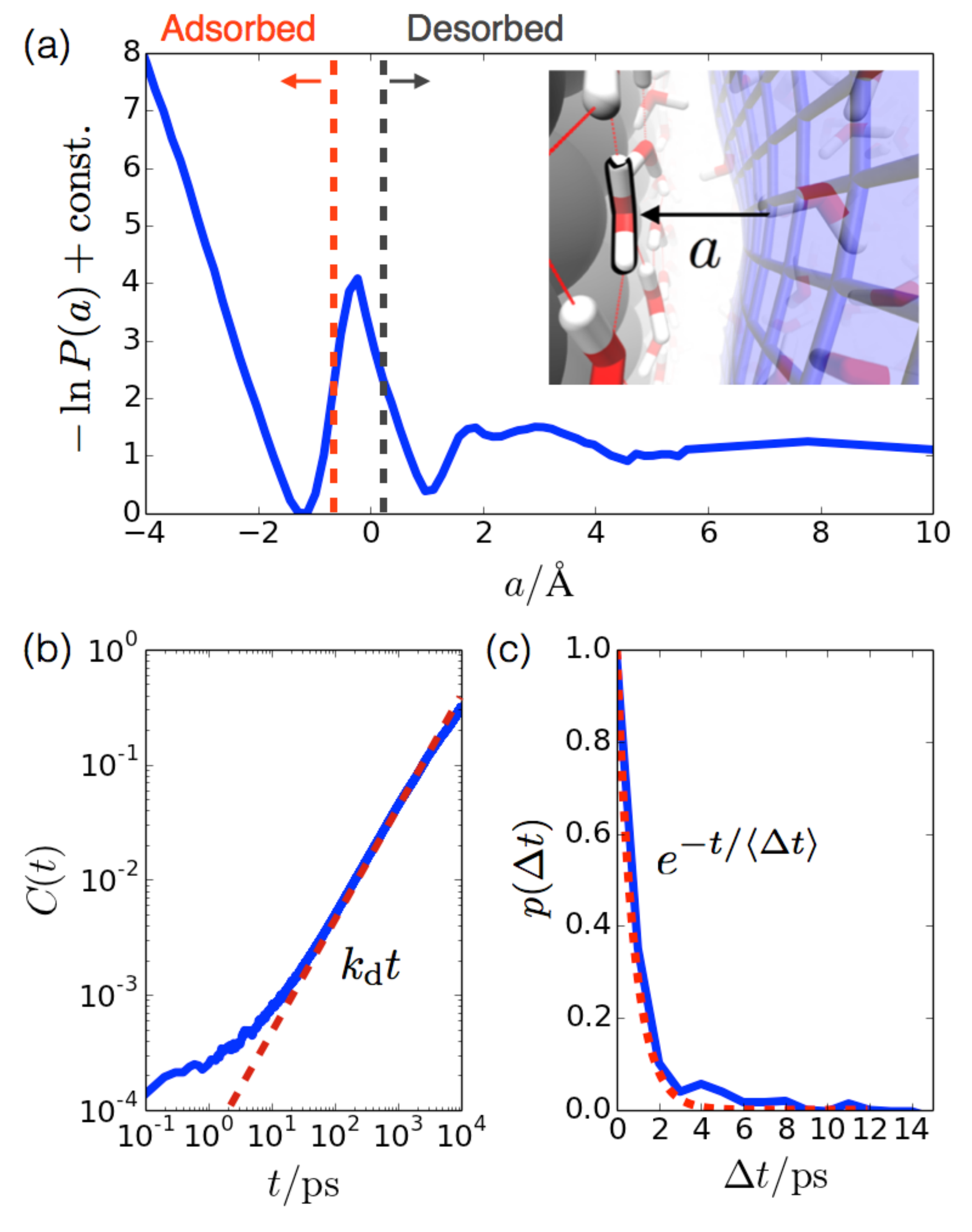}
\caption{Distributions and timescales for desorption. (a) Potential of mean force for moving a water molecule relative to the instantaneous interface. The red and black dashed lines denote the maximum and minimum values for defining basins of adsorbed and desorbed states.  $\ln P(a)$ is shifted by a constant so as to make its minimum zero.  (inset) Characteristic snapshot of an adsorbed molecule and the instantaneous liquid interface. (b) Flux correlation function, blue solid line, as defined in Eq.~\ref{Eq:flux}. The dashed red line is a linear fit to $C(t)$ whose proportionality constant yields the rate for desorption. (c) Distribution of instanton times for the desorption event, blue, as defined in Eq.~\ref{Eq:instanton}, and a reference exponential decay with characteristic time $\Delta t=3$ps.}
\label{Fi:2}
\end{center} 
\end{figure}

The instantaneous liquid interface, $\mathbf{s}$, is identified implicitly~\cite{Willard:2010p1954}: 
\begin{equation}
\tilde{\rho}(\mathbf{s}) = \rho_\ell/2 \, ,
\end{equation}
where $\rho_\ell$ is the bulk liquid density, and $\tilde{\rho}(\mathbf{r})$ is the molecular density coarse grained over a length scale $\xi$, i.e.,
\begin{eqnarray}
\label{eq:rho}
\tilde{\rho}(\mathbf{r}) &=& \sum_i^N \phi(\mathbf{r}-\mathbf{r}_i; \xi) \, .
\end{eqnarray}
Here, $\mathbf{r}_i$ is the position of the oxygen of the $i$th water molecule, there are $N$ molecules in total, and $\phi(\mathbf{r})$ is chosen to be a spherically symmetric and normalized Gaussian function, truncated and shifted to become zero continuously beyond at $|\mathbf{r}|=3\xi$.  We take $\xi=2.5 \mathrm{\AA}$. 

The position of the $i$th molecule relative to the interface is 
\begin{equation}\label{Eq:a}
a_i = \mathbf{n(\mathbf{s}_{\mathbf{r}_i})} \cdot (\mathbf{r}_i -\mathbf{s}_{\mathbf{r}_i})\,,
\end{equation}
where $\mathbf{n}(\mathbf{s})$ is the unit vector normal to the interface at $\mathbf{s}$ pointing away from the bulk, and $\mathbf{s}_\mathbf{r}$ is the point on $\mathbf{s}$ closest to $\mathbf{r}$.  See inset to Fig.~\ref{Fi:2}.  $a_i$ is a scalar that can be positive or negative. We employ the convention that omits the tagged molecule from the density used to define the instantaneous interface.  The alternative, including the tagged molecule in the density, gives behaviors qualitatively similar to those shown below.  

With an umbrella sampling procedure developed previously~\cite{varilly2013water}, we calculate $P(a) = \langle \delta(a - a_i) \rangle$, where the angle brackets, $\langle \cdots \rangle$, denote equilibrium average.  The function $-\ln P(a)$ is the associated potential of mean force in units of Boltzmann's constant times temperature, $\kB T$.  This function is shown in Fig.~\ref{Fi:2}(a) for water adjacent to the 100 Pt surface. It has three notable features: (1) For $a<0$, there is a narrow minima reflective of the bound adlayer.  (2) Between the bound adlayer and the bulk liquid, there is a free energy barrier peaked at $a\approx 0$ with a height relative to the surface layer of $4 k_\mathrm{B}T$.  (3) The bulk liquid, $a>2$\,\AA, has a more diffuse basin than that of the bound layer. Similar qualitative features appear in the analogous free energy as a function of absolute position shown in the Appendix.

In view of these features, we define the following dynamic indicator functions for a tagged molecule, $i$, belonging to the adsorbed state, $h_i$, or desorbed state, $g_i$. These are defined as
\begin{eqnarray}\label{Eq:ha}
h_i&=& 1, \, \, \, \mathrm{if} \, \, \, a_i \le -0.7\, \mathrm{\AA}\\ \notag
&=& 0, \, \, \, \mathrm{else}
\end{eqnarray}
and 
\begin{eqnarray}\label{Eq:hb}
g_i&=& 1, \, \, \, \mathrm{if} \, \, \, a_i \ge 0.3\, \mathrm{\AA}\\ 
&=& 0, \, \, \, \mathrm{else} \notag
\end{eqnarray}
and are indicated in Fig.~\ref{Fi:2}(a). The time-correlation function between the two
\begin{equation}\label{Eq:flux}
C(t) = \frac{\langle h_i \, g_i(t)\rangle}{\langle h_i  \rangle } \, ,
\end{equation}
is shown in Fig.~\ref{Fi:2}(b).  We use standard notation for the time correlation function, i.e., the equilibrium average is over initial conditions, and $h_i(t)$ is the desorbed state indicator function evaluated for the configuration at time $t$.  The correlation function is independent of the particle label, $i$, because all the water molecules are indistinguishable at equilibrium.  Later, we generalize to nonequilibrium cases where the particle-label index is important.

This correlation function is estimated with straightforward molecular dynamics simulations by harvesting 2000 trajectories bridging adsorbed and desorbed basins (see Appendix for details). The short-time behavior of $C(t)$ is indicative of transient barrier crossings and recrossings and the longer-time behavior exhibits the linear growth expected for a rate process\cite{chandler1987introduction}.  The time derivative, $d C(t)/dt $, reaches its plateau value after a transient time, and this transient time is approximately the mean instanton time, $\langle \Delta t \rangle$, where a formula for the instanton time of a specific trajectory, $\Delta t$, is given below. With the basins as defined above (Eqs.~\ref{Eq:ha} and \ref{Eq:hb}), this time is about 3 ps.  Far beyond that time (i.e. $t \gg \langle \Delta t \rangle$), the slope determines the rate constant, i.e., $C(t) \sim k_\mathrm{d} \, t$.  The best fit to the data in Fig.~\ref{Fi:1}(b) yields $k_\mathrm{d} \approx 2.6 \times10^{-2} \,\mathrm{ns}^{-1}$. In other words, given the surface density of water on the electrode, one desorption event occurs roughly every 3 ns per nm$^2$.

The rate for desorption is independent of the basin definitions, but the typical time to reach the plateau, $\langle \Delta t \rangle$, is sensitive to these definitions. This sensitivity arises from the possibility of recrossing events as the molecule lingers near the top of a free energy barrier. As a quantitative measure of the time it takes to cross the barrier, for each single trajectory of duration $t_\mathrm{obs}$ in which the tagged molecule $i$ crosses from adsorbed state to desorbed state, we compute the instanton time,
\begin{eqnarray}\label{Eq:instanton}
\Delta t &=& \int_0^{t_\mathrm{obs}}\,\mathrm{d}t\, \left[ 1-h_i(t) \right]\,\left[1-g_i(t) \right] \, .
\end{eqnarray}
From an ensemble of those trajectories, we obtain the distribution of instanton times, which is plotted in Fig.~\ref{Fi:2}(c).  The distribution is roughly exponential, consistent with different desorption events being uncorrelated.  The time constant for the exponential is about 3 ps. 

This average value,  $\langle \Delta t \rangle$, is in line with the typical time for a water molecule to diffuse $1\mathrm{\AA}$ in the homogeneous liquid, which is about 5 ps.  On the other hand, a much larger instanton time is found when using definitions of $h_i$ and $g_i$ based upon the separation from the metal surface or mean interface, rather than the separation from the instantaneous interface.  That alternative separation is given by $z_i$, the component of the vector $\mathbf{r}_i$ perpendicular to the plane of the static surfaces.  In that case, replacing $a_i$ with $z_i$ in Eqs.~\ref{Eq:ha} and \ref{Eq:hb}, we find that the average instanton time increases to 20 ps.  (See Appendix B.) This larger time indicates the increased likelihood for recrossing events along this coordinate. Indeed this longer timescale is commensurate with the average time for height fluctuations of the instantaneous interface~\cite{Willard:2010p1954,limmer2013hydration}, suggesting that the physical origin for the additional recrossing events lie in the relaxation of these interfacial fluctuations.

\begin{figure}[bh]
\begin{center}
 \includegraphics[width=8.5cm]{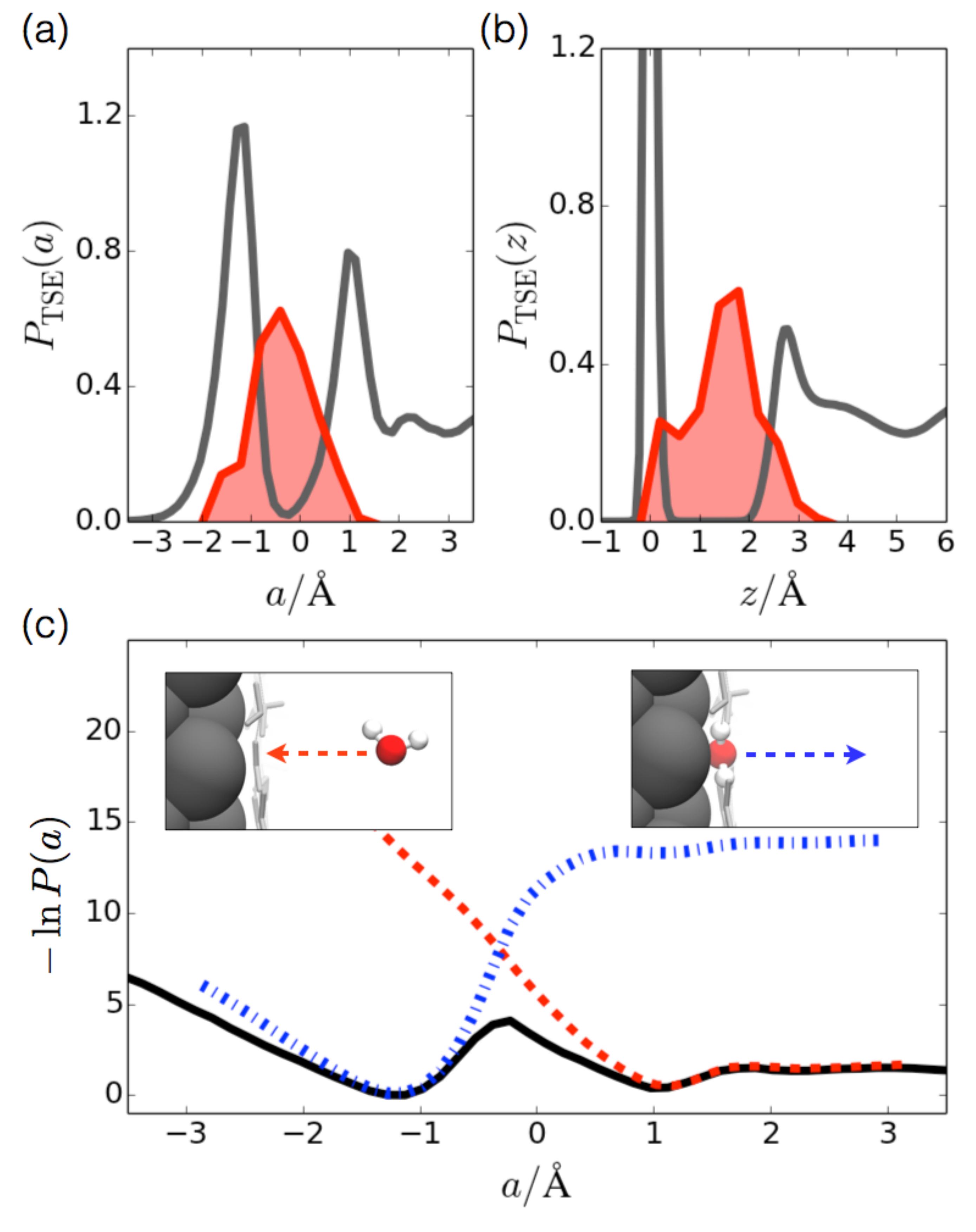}
\caption{ (a) The probability distribution for the relative distance to the instantaneous interface, for a tagged molecule within the transition state ensemble. In grey is the equilibrium distribution. (b) The probability distributions for the absolute position for a tagged molecule within the transition state ensemble.  In grey is the equilibrium distribution. (c) Three free energy profiles relative to the instantaneous interfaces (in units of $\kB T$): for moving a water molecule reversibly (black);  for moving a water molecules either from the surface or from the bulk, reversibly with respect to the bulk waters, but irreversibly with respect to the other surface adlayer molecules, which are fixed in one typical configuration (blue or red, respectively).}
\label{Fi:3}
\end{center} 
\end{figure}

\section{Analyzing the transition state ensemble}

The variable $a_i$ -- the separation of a water molecule from the instantaneous interface -- serves as a distinguishing order parameter for adsorbed and desorbed states, but it is not sufficient to characterize pathways between adsorbed and desorbed states. This fact is demonstrated now through analysis of the transition state ensemble (TSE).  

\begin{figure*}[ft]
\begin{center}
 \includegraphics[width=17cm]{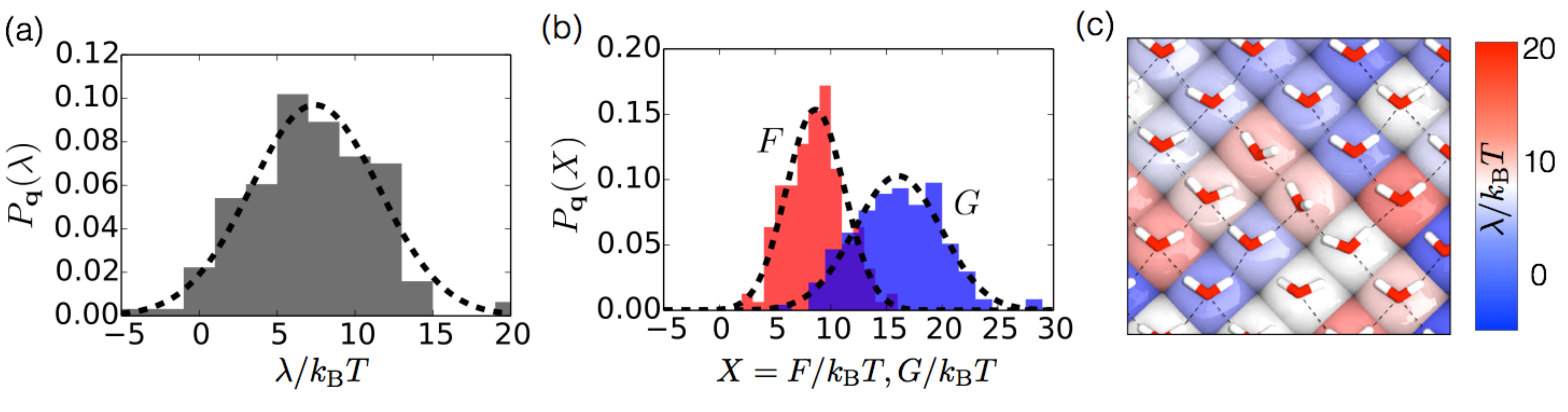}
\caption{(a) Probability distribution of surface reorganization energies, $\lambda$ for a given surface configuration $\mathbf{q}$. The dashed line is a Gaussian fit. (b) Probability distributions of the work associated with pulling an adsorbed molecule off of the electrode, with surface configuration $\mathbf{q}$, and under different constraints. Dashed lines are Gaussian fits. (c) Characteristic snapshot of adsorbed waters, with metal atoms colored by their value of surface reorganization energy, $\lambda$.}
\label{Fi:4}
\end{center} 
\end{figure*}

\subsection{Committor and definition of TSE}
Members of the TSE can be harvested from the trajectories used to compute the correlation function, $C(t)$. These are the configurations from which short trajectories commit to either reactant or product with equal likelihood~\cite{bolhuis2002transition}. In particular, for trajectories that begin at coordinate $\mathbf{x} = \{\mathbf{r}_1,...,\mathbf{r}_N\}$ with randomly chosen velocities and propagate for a time $t_\mathrm{obs}$, the commitment probability, $p(\mathbf{x},t_\mathrm{obs})$, is the fraction of trajectories that end up in the desorbed basin.  Here, the trajectory time is larger than the transient time but not so large to permit multiple transitions, i.e. $\langle \Delta t \rangle \lesssim t_\mathrm{obs} \ll 1/k_\mathrm{d}$, and accordingly, the TSE is the subset of configurations with $p(\mathbf{x},t_\mathrm{obs})=1/2$.  To estimate $p(\mathbf{x},t_\mathrm{obs})$, we use $t_\mathrm{obs}=10$ ps and average over 10 separate realizations of the random velocities taken from the Maxwell-Boltzmann distribution. If this estimate falls within a 90$\%$ confidence interval of 0.5, we say $\mathbf{x}$ is a member of the TSE. 

The distribution of these configurations, $P_\mathrm{TSE}(\mathbf{x})$, is difficult to interpret because $\mathbf{x}$ is a high-dimensional variable.  To make headway, we project onto a low-dimensional representation.  The notation we use for the resulting marginal distributions is
\begin{equation}
P_\mathrm{TSE}(A) = \int d \mathbf{x} \, P_\mathrm{TSE}(\mathbf{x}) \, \delta \left( \hat{A}(\mathbf{x})-A \right),
\end{equation}
where $\hat{A}(\mathbf{x})$ is the dynamical variable, and $A$ is its observed value. 

Figure \ref{Fi:3}(a) refers to the TSE projected onto the coordinate $a_i$, i.e., the separation of the desorbed molecule from the instantaneous bulk liquid interface (see Eq.~\ref{Eq:a}). Figure \ref{Fi:3}(b) refers to the TSE projected onto the coordinate $z_i$, i.e., the separation from the average position of the adlayer.  Both distributions are peaked in the interfacial region between the bound adlayer and the bulk liquid.  The relative widths of the distribution are $\langle |\delta a | \rangle /\langle a \rangle \approx 0.2$ and $ \langle |\delta z | \rangle /\langle z \rangle \approx 0.35$. By this metric, reference to the instantaneous interface (i.e., the displacement $a_i$), is a better descriptor of the TSE than reference to the mean interface (i.e., the displacement $z_i$). However, in both cases, Figs. \ref{Fi:3}(a) and (b) show, the transition states still occur over a significant range of separations. These distribution widths indicate that the desorption reaction coordinate has important components orthogonal to both of these distances. 

\subsection{Role of adlayer reorganization}

Important variables missing from $a_i$ (or $z_i$) involve configurations of the water molecules bound to the metal surface. We use the symbol $\mathbf{q}$ to stand for the collection of coordinates that together with $\mathbf{r}_i$ specify the configuration of that adlayer.  $\mathbf{q}$ is significant because each desorption event is accompanied, at least transiently, by the creation and relaxation of a surface vacancy.  Effects of this reorganization can be seen by juxtaposing the reversible work function, $-\ln P(a)$, with two nonequilibrium counterparts that depend upon $\mathbf{q}$.  

The first counterpart is 
\begin{equation}
G_{i,\mathbf{q}}(a, a_0) =  \frac{1}{\kB T} \int_{a_0}^{a} \mathrm{d}a' \langle f_i(a')  \rangle_\mathbf{q}\,.
\end{equation}
Here, $f_i(a)$ is the instantaneous force on $a_i$ evaluated at $a_i = a$ [i.e., $f_i(a) = - (\partial U/\partial a_i ) |_{a_i = a}$, where $U=U(\mathbf{x})$ is the total potential energy of the system], and $\langle \cdots  \rangle_\mathbf{q}$ stands for equilibrium average with $\mathbf{q}$ fixed.  $G_{i,\mathbf{q}}(a, a_0)$ is therefore the reversible work (in units of $\kB T$) to move $a_i$ from a reference position, $a_0$, to $a$ with the rest of the adlayer fixed in configuration $\mathbf{q}$.  Averaging this constrained work over the equilibrium distribution of $\mathbf{q}$ yields $-\ln P(a)$ to within an additive constant.  Without that averaging, the constrained free energy function is generally different from the reversible work function.  Two typical examples of the constrained free energy function are illustrated in Fig.~\ref{Fi:3}(c) with $a_0 = -1$\,\AA \, (the dashed blue line) and with $a_0 = 1$\,\AA \, (the dashed red line). The respective $\mathbf{q}$'s are taken from the equilibrium ensembles of configurations with $\mathbf{r}_i$ set at a value where $a_i = a_0$.  Comparison with $-\ln P(a)$ (the black line) shows that adlayer reorganization is indeed a significant and irreversible effect.

Some of that reorganization occurs on relatively short time scales, which brings us to the second nonequilibrium counterpart,
\begin{equation}
F_{i,\mathbf{q}}(a, a_0;t) =  \frac{1}{\kB T} \int_{a_0}^{a} \mathrm{d}a' \langle f_i(a')  \rangle_{\mathbf{q},t}\,.
\end{equation}
Here, $\langle \cdots \rangle_{\mathbf{q},t}$ stands for the time average over a trajectory of length $t$, with $\mathbf{q}$ as the \emph{initial} configuration for the rest of the adlayer.  $\mathbf{q}$, in this case is not constrained in the subsequent dynamics.  $F_{i,\mathbf{q}}(a, a_0; t)$ is then the average work done to move $a_i$ from $a_0$ to $a$ in a time $t$ starting with the rest of the adlayer in configuration $\mathbf{q}$.  As $t \rightarrow \infty$, this time-dependent work function becomes $-\ln P(a)$ to within an additive constant.  

On the way to that limiting behavior, there are two relaxation regimes.  The first coincides with reorganization directly spurred by the desorption event.    It takes place on times scales between 10 and 100\,ps.  The second coincides with equilibration of the adlayer's transient heterogeneity.  It takes place on time scales of 1 to 100\,ns. Thus, $F_{i,\mathbf{q}}(a, a_0;t)$ for 1\,ns $<t<$ 10\,ns gives a measure of relaxation for one realization of the transient heterogeneity, and as such, it has a distribution of values characteristic of the distribution of transient heterogeneity.

In general, for fixed $a$ and $a_\mathrm{o}$, $G_{i,\mathbf{q}}$ and $F_{i,\mathbf{q}}$ are different due to the different constraints on the relaxation of surface bound molecules.  When $G_{i,\mathbf{q}}$ and $F_{i,\mathbf{q}}$ are computed with $a_\mathrm{o}$ near the equilibrium position of adsorbed state and $a$ in the bulk, their deviation signals the additional local free energy that must be dissipated in order to accommodate a molecule leaving the surface. In this way, by locating tagged regions where both measures are equivalent,  we can isolate regions on the surface where reorganization is facile and admitting to desorption events.  Specifically, we can define the deviation between these two work functions for moving a tagged molecule between the surface and bulk for a given initial surface configuration
\begin{equation}
\lambda = G - F \, .
\end{equation}
We adopt the shorthand, $G=G_{i,\mathbf{q}}(a,a_0)$ and $F=F_{i,\mathbf{q}}(a,a_0)$ where each is evaluated with $a_0 = -1$ and $a = 3$. To determine the reorganization energy dissipated through local surface relaxation, our results are not sensitive to these precise values as long as $a_0$ is near the maximum of $P(a)$ and $a$ is within the large $a$ plateau. A histogram of $\lambda$, denoted $P_{\mathbf{q}}(\lambda)$ to emphasize that each value of $\lambda$ depends on the initial surface configuration, is shown in Fig.~\ref{Fi:4}a for the 100 surface. This is determined from the histograms of $G$ and $F$ shown in Fig.~\ref{Fi:4}b. Within our limited statistics each distribution is Gaussian. The peak value of $G$ is larger than $F$ by 7$\kB T$ and its distribution is wider by a factor of 1.5. The larger average value of $G$ over $F$ is a consequence of the additional constraint on the surface. The variance in $\lambda$ is not the sum of the variances of $G$ and $F$, implying some degree of correlation, however this correlation is small, with a correlation coefficient of 0.2. Distributions of $\lambda, G$ and $F$ for the 111 surface are shown in the Appendix. 

For the 100 surface, we find that on average the local number of persistent hydrogen bonds made adjacent to a tagged site increases with increasing $\lambda$. We define a persistent hydrogen bond as an oxygen oxygen distance of less than $\mid \mathbf{r}_\mathrm{oo}\mid= 3.5\mathrm{\AA}$ and a OH bond angle relative to $\mathbf{r}_\mathrm{oo}$ less than 30$^o$ with positions coarse-grained over 1 ps. This is a reflection of the stable monolayer surface structure, which is commensurate with the 4-fold coordination of the exposed lattice and water's preferred hydrogen bonding patterns. Defects that disrupt this four-coordinated network allow for increased mobility of the water molecules orientation\cite{willard2013characterizing}, and consequently these sites have a lower reorganization energy. However, there is not a perfect correlation between persistent hydrogen bonds and reorganization energy, the correlation coefficient is 0.5. This low correlation is because the reorganization energy is not a local quantity, and rather can reflect more subtle strains in the hydrogen bonding network that add up over neighboring molecules and allow for facile relaxation.  Hydrogen bonds drawn in Fig.~\ref{Fi:4}(c) are the persistent hydrogen bonds discussed above. The trend relating hydrogen bonding and $\lambda$ can be observed from this snapshot, though also shown are exceptions such as molecules with only 3 persistent hydrogen bonds that can still have relatively high values of $\lambda$. While many-bodied in nature, as illustrated in Fig.~\ref{Fi:4}(c) the spatial distribution of reorganization energies are nevertheless made up of uncorrelated domains that vary over the surface with negligible correlation lengths, similar to that found for the induced charge distributions on the metal centers\cite{limmer2013charge}.

\begin{figure}
\begin{center}
 \includegraphics[width=8.5cm]{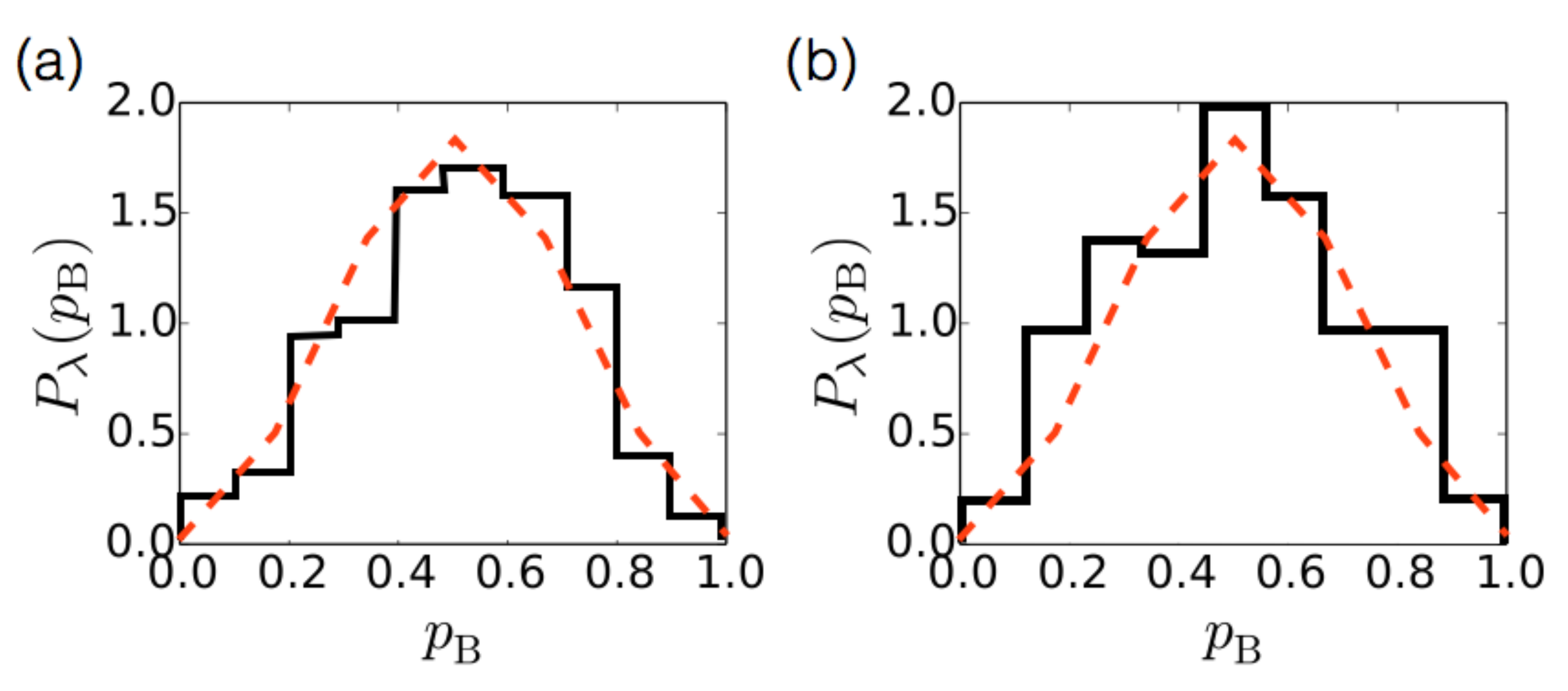}
\caption{Commitor distribution function averaged over an ensemble constrained to $\lambda\approx 0$ for the (a) 100 and (b) 111 surfaces. The red dashed lines are Bernoulli distributions for 11 trials.}
\label{Fi:5}
\end{center} 
\end{figure}
\begin{figure*}
\begin{center}
 \includegraphics[width=17cm]{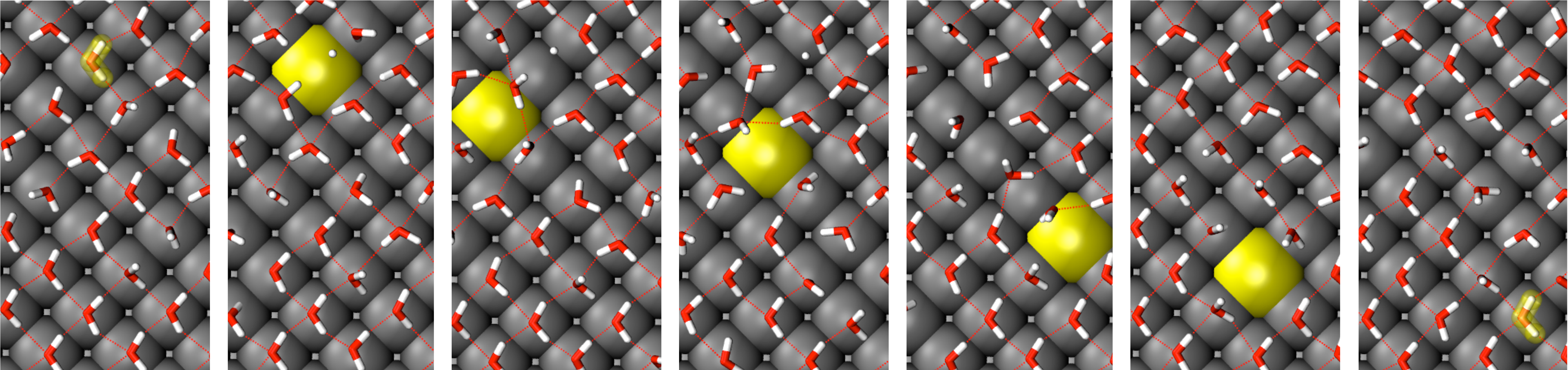}
\caption{Characteristic time series for vacancy diffusion during an exchange process. The exchange pair are highlighted in yellow as is the surface vacancy. Snapshots are approximately 5 ps apart. A movie of this trajectory found here: \small{https://youtu.be/P8LfvycgdpI}.}
\label{Fi:6}
\end{center} 
\end{figure*}
As an ultimate test of the relevance of this surface coordinate for the reaction dynamics, we have calculated the distribution of commitment probabilities over an ensemble initialized with $\lambda \approx 0$. This type of test, first proposed by Geissler, et. al\cite{geissler1999kinetic} distinguishes potential reaction coordinates that well characterize the TSE only on average due to a near equivalency of populations of configurations committed solely to the A or B basins. Figure \ref{Fi:5} plots the distribution of commitor values for fixed $\lambda$, $P_\lambda(p_\mathrm{B})$ with a reference to a binomial distribution with 11 trials, corresponding to the mean number of trials used in the evaluation of the commitment probability. The peak near 0.5 confirms that these constraints impose a proximity to the true seperatrix, and their width is consistent with a finite number of trials. Figure \ref{Fi:5}  plots these distributions for both the 100 and 111 surfaces of platinum, illustrating the degree of generality of this reaction coordinate. Details on the calculations for the 111 surface are provided in the Appendix. 

\section{Correlations between exchange pair}
Most of this work has focused on the microscopic events leading a water molecule to desorb from the electrode surface. By microscopic reversibility, the pathways for the adsorption of a water molecule from the bulk onto an electrode must be the time-reversed as those for the desorption of a water molecule from the electrode into the bulk. Moreover, from the law of mass action,  the rate of adsorption must be equal to the rate for desorption times the free energy difference between the two states. Therefore, it would seem that by studying the forward process of desorption, all relevant information of exchange can be extracted.  Though this is mostly true, neither of the two previous statements prohibits the existence of dynamic correlations between the exchange pair. 

The extent of correlations between the exchange pair is evident in Fig.~\ref{Fi:6}, which depicts a characteristic time series of the diffusion of the vacancy left behind after a desorption event. The $k$th vacancy is created each time a desorption event occurs, at the position $\mathbf{r}$, the location of the closest platinum atom to the desorbing molecule, is denoted $n_k(\mathbf{r},t)$ and carries a numerical value of 1.
Similarly, a vacancy is annihilated each time an adsorption event occurs, which changes the numerical value of $n_k(\mathbf{r},t)$ to 0.
A typical vacancy is highlighted in yellow in Fig.~\ref{Fi:6} and is shown to diffuse through the surface over the course of 30 ps before being annihilated by an adsorbing molecule. The diffusion of the vacancy requires the hydrogen bonding network of the adlayer to reorganize, which is does following a process similar to that described by Ref.~\onlinecite{ranea2004water} in low water coverage \emph{ab initio} simulations on the palladium 111 surface, whereby a hydrogen bonded pair is anchored by one water on a metal center and rotates its hydrogen bonded partner that is weakly bound to the surface between two adjacent sites. 

\begin{figure}[b]
\begin{center}
 \includegraphics[width=8.5cm]{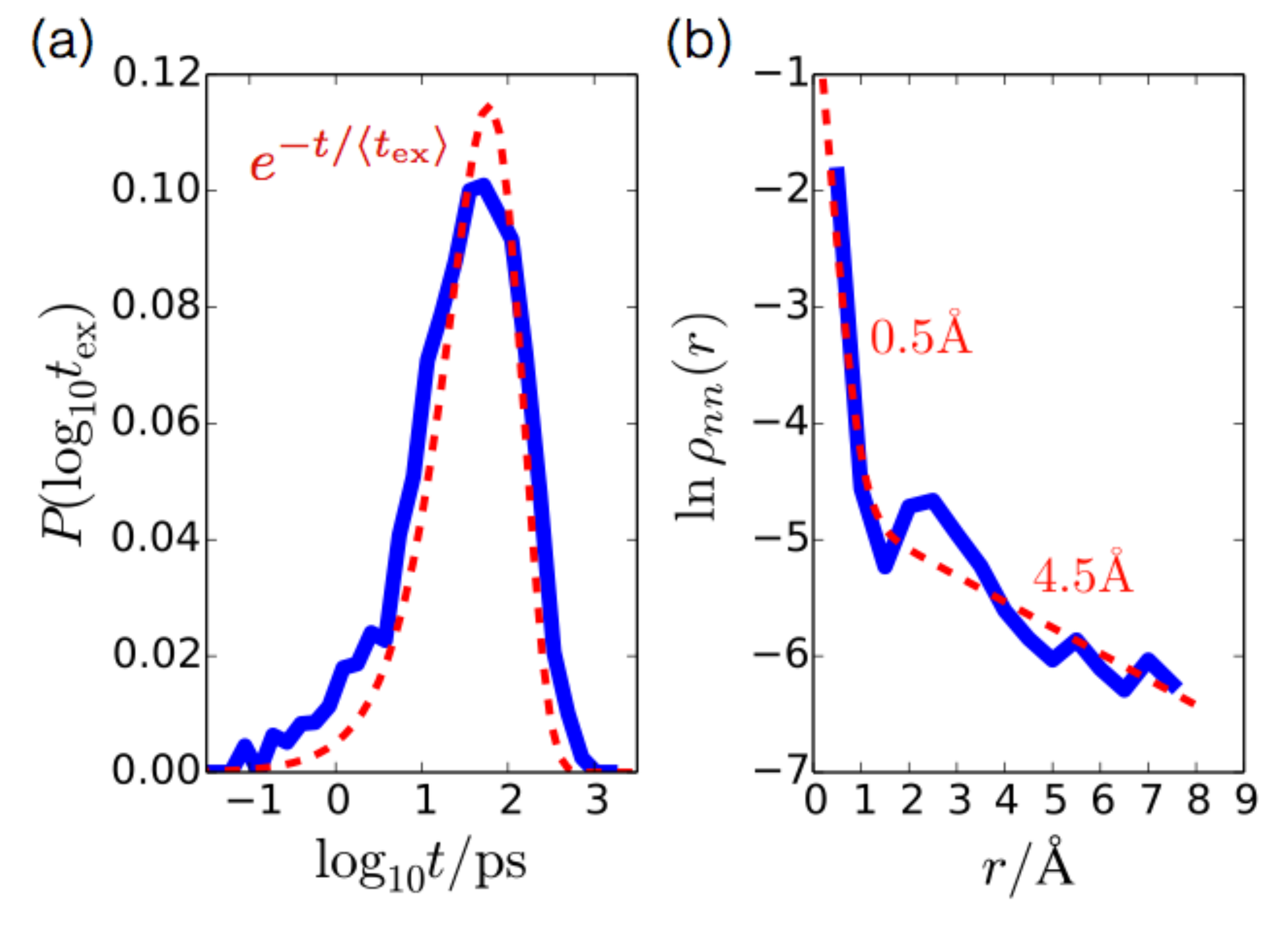}
\caption{(a) Probability distribution of lifetimes for a surface vacancy. The mean lifetime is $\langle  t_\mathrm{ex}  \rangle = $ 60 ps. (b) Probability distribution for the distance between the surface sites for vacancy creation and the following site of vacancy annihilation. The dashed red line represents an exponential distribution of exchange times, which on the the logarithmic scale has a characteristic single peak at its mean value. Characteristic decay lengths are 0.5$\mathrm{\AA}$ and 4.5$\mathrm{\AA}$.}
\label{Fi:7}
\end{center} 
\end{figure}

The lifetime of a surface vacancy determines the conditional waiting time for an adsorption event, given a desorption event just occurred, in a way that is local and not obscured in the limit of large surfaces. We define the $t_\mathrm{ex}$, as the lifetime for the $k$th vacancy,
\begin{equation}
t_\mathrm{ex} = \int \mathrm{d}t\, n_k(\mathbf{r},t) \, .
\end{equation}
Figure \ref{Fi:7}(a) shows the distribution of lifetimes in logscale, which is strongly peaked at, $\langle  t_\mathrm{ex}  \rangle=$60 ps. This time is much less then the mean time for adsorption, which means most adsorption events result from correlated surface configurations that transport localized defects from one place on the surface to another. 

The pair correlations for the location of the creation and annihilation of the surface vacancy mirror the vacancy lifetimes. Specifically, we define a pair correlation function for the location of the vacancy creation and it subsequent annihilation,
\begin{equation}
\rho_{nn}(|\mathbf{r}-\mathbf{r}'|) =\frac{\langle n_k(\mathbf{r},t) n_k(\mathbf{r}',t+t_\mathrm{ex}) \rangle}{\langle n_k(\mathbf{r}',t+t_\mathrm{ex}) \rangle}  \, ,
\end{equation}
which due to rotational invariance depends only on the magnitude of $\mathbf{r}-\mathbf{r}'$, and due to time translational invariance, depends only on $t_\mathrm{ex}$. This correlation function are shown in Fig.~\ref{Fi:7}(b). While it is most probable that the vacancy does not diffuse, the decay in Fig.~\ref{Fi:7}(b) is slow, and there is significant probability that vacancies diffuse over nanometer lengths. This function can be well fit with a biexponential decay with characteristic lengths of 0.5$\mathrm{\AA}$ and 4.5$\mathrm{\AA}$. The long waiting time and large distance correlations can be understood from what has already been found regarding the kinetics of desorption. Namely, adsorption occurs due to a localized defect in the hydrogen bond network of the surface waters coming into contact with a rare, fleeting fluctuation of the interfacial density. 

\begin{acknowledgments} 
Early work on this project in its early stages was supported by the Helios Solar Energy Research Center of the U.S. Department of Energy under Contract No. DE-AC02-05CH11231. In its final stages, it was supported by the Director, Office of Science, Office of Basic Energy Sciences, Materials Sciences and Engineering Division and Chemical Sciences, Geosciences, and Biosciences Division under the same DOE contract number. DTL is now supported as a fellow of the Princeton Center for Theoretical Science. This research used resources of the National Energy Research Scientific Computing Center, a DOE Office of Science User Facility supported by the Office of Science of the U.S. Department of Energy under Contract No. DE-AC02-05CH11231.
\end{acknowledgments}

\appendix
\section{Methods}
The model we adopt to describe the interface utilizes a standard classical force field to compute the intermolecular interactions between water molecules, namely the SPC/E potential\cite{Berendsen:1987p8660}, and a semi-empirical description of an ideally polarizable metal surface held at constant electrical potential developed by Seipman and Sprik\cite{Siepmann:1995p4868,reed2008electrochemical}. The atomic geometry of the electrode, along with the details of the water-metal interactions are consistent with that of crystalline platinum. The water-metal interaction is based on simple quantum-chemical calculations and reproduces the large binding energy and ground state geometry of a single water molecule\cite{Siepmann:1995p4868}. Classical contributions to the potential of zero charge\cite{limmer2013hydration} and capacitance\cite{Willard:2008p8256,limmer2013charge} have been shown to be in good agreement with more detailed models. 

The system dimensions and particle numbers are the same as in previous work. Specifically, a 3.5nm by 3.5nm by 5nm simulation box is filled with nearly 2000 water molecules and close to 1000 platinum atoms in 2 slabs, each three rows wide, resulting in a mean bulk water density in the center of the slab of 1 g/cc. Equilibration is difficult due to the slow surface water dynamics, so an ensemble of initial conditions is generated by annealing the system at 400 K over 1 ns, and cooled to 298 K over another ns. Production runs used to generate the ensemble of reactive trajectories are 20 ns, which for each 20 surfaces yield approximately 2000 desorption events. The integrator uses a langevin thermostat, with characteristic time constant of 1ps, and molecular constraints imposed with SETTLE\cite{miyamoto1992settle}. Long ranged electrostatics are accounted for using the appropriate Ewald summation for mixed point and Gaussian charge distributions\cite{gingrich2010ewald}, and the two-dimensional periodicity is approximated with the standard slab correction\cite{yeh1999ewald}, which elongates the simulation cell by a factor of 5. We use the same methodology for both 100 and 111 surfaces. Calculations are done with a modified version of the LAMMPS code\cite{Plimpton:1995p3851}.

Umbrella sampling with MBAR\cite{shirts2008statistically} is used to estimate free energies in Figs.~\ref{Fi:2}-\ref{Fi:4}. For free energy calculations with quenched surface configurations, production runs of 1 ns are sufficient to converge local distributions, while for calculations with evolving surface configurations production runs of 10 ns are required. In each case, 12 windows are equally spaced throughout the $a$ coordinate and harmonic biasing potentials with spring constants of 8.0 kcal/mol $\mathrm{\AA}^2$ are used. 
\begin{figure}[t]
\begin{center}
 \includegraphics[width=8.5cm]{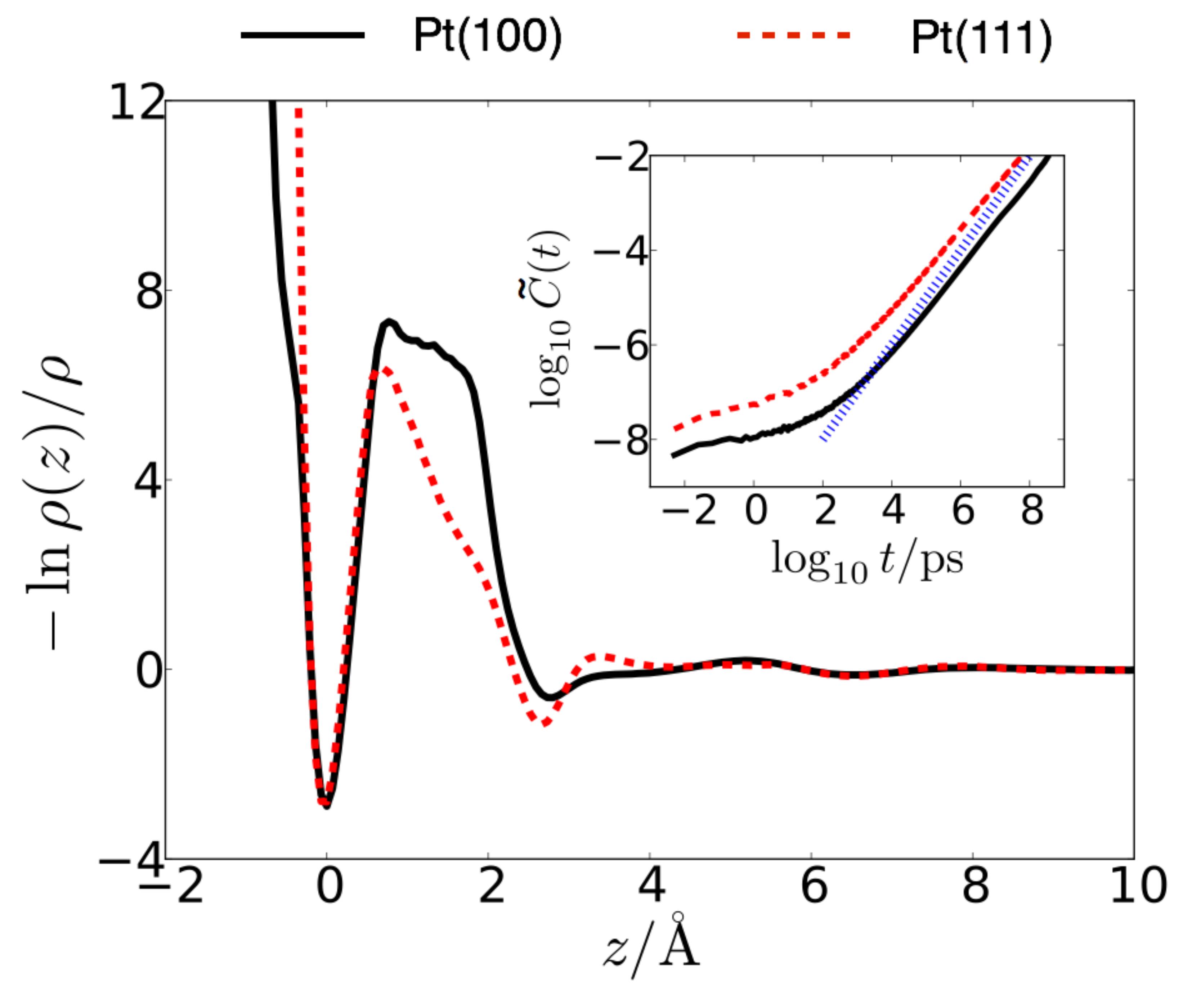}
\caption{Potential of mean force for moving a water molecule relative to the absolute distance from the electrode surface. (inset) Flux correlation functions for $z$ basin definitions for the 100, red, and 111, black surfaces. In blue is a line with unit slope.}
\label{Fi:8}
\end{center} 
\end{figure}

\section{Cartesian coordinate basin definition}

An alternative means for defining the basin definitions in Eq.~\ref{Eq:ha} and \ref{Eq:hb} would be to use the absolute position away from the electrode, rather then the position relative to the instantaneous liquid interface. To construct the analogous indicator functions it is first useful to calculate the free energy for moving a molecule in along this coordinate. This free energy is easily calculated by taking the negative logarithm of the average water density as a function of $z$, or $\rho(z)$.  

Figure \ref{Fi:8} plots the free energy as a function of $z$ for both the 100 and 111 surface. Similar to the results in Fig.~\ref{Fi:1}(a), both curves exhibit a narrow basin near the electrode, $z<1\mathrm{\AA}$, and a large barrier between this bound water layer and the diffuse bulk liquid, for $z>3\mathrm{\AA}$. The barrier heights for the 100 and 111 surfaces are both large, at 9 and 10, respectively. These barriers are much larger then that depicted in Fig.~\ref{Fi:1}(a). This is a reflection of the reduced force needed for collective fluctuations to reorganize around the tagged molecule, rather than the forces to move a molecule relative to a static environment. 

We define indicator functions analogously as in the main text but now as a function of $z$, 
\begin{eqnarray}\label{Eq:ha1}
\tilde{h}_i &=& 1 \, \, \, \mathrm{if} \, \, \, z_i \le 0.8\, \mathrm{\AA}\\
&=& 0 \, \, \, \mathrm{else}\notag
\end{eqnarray}
and 
\begin{eqnarray}\label{Eq:hb1}
\tilde{g}_i &=& 1 \, \, \, \mathrm{if} \, \, \, z_i \ge 3\, \mathrm{\AA}\\
&=& 0 \, \, \, \mathrm{else} \notag
\end{eqnarray}
and from these indicator functions we similarly compute a flux correlation function
\begin{eqnarray}
\tilde{C}(t) &=& \frac{\langle \tilde{h}_i \tilde{g}_i(t) \rangle}{\langle \tilde{h}_i  \rangle } \, ,
\end{eqnarray}
which is plotted in the inset to Fig.~\ref{Fi:8} for both surfaces. In harmony with the result in the main text the asymptotic behavior of $\tilde{C}(t)$ is $\tilde{k}_d t$, where $\tilde{k}_d$ is the rate constant using these basin definitions. As before this is exacted by a best fit line and results in $\tilde{k}_d=0.053\,\mathrm{ns}^{-1}$ for the 100 surface and $\tilde{k}_d=0.028\,\mathrm{ns}^{-1}$ for the 111 surface. 

As alluded to in the Section \ref{Sec:Time}, while use $\tilde{h}_i$ and $\tilde{g}_i$ rather then $h_i$ and $g_i$ does not change the overall rate constant, it does change the time for reaching steady-state. For the 100 surface this is significant, changing the lag time from it from 3ps to 20 ps. For the 111 surface it is not as significant, changing the mean time from 3ps to 12ps. This behavior is evident in the inset to Fig.~\ref{Fi:8}, where the linear growth regime can be gauged by the line of unit slope also plotted. The origin of this difference between the two surfaces is due to the 100 being significantly more hydrophobic than the 111 surface\cite{limmer2013hydration}, which thus enables much larger interfacial density fluctuations.

\begin{figure}[t]
\begin{center}
 \includegraphics[width=8.5cm]{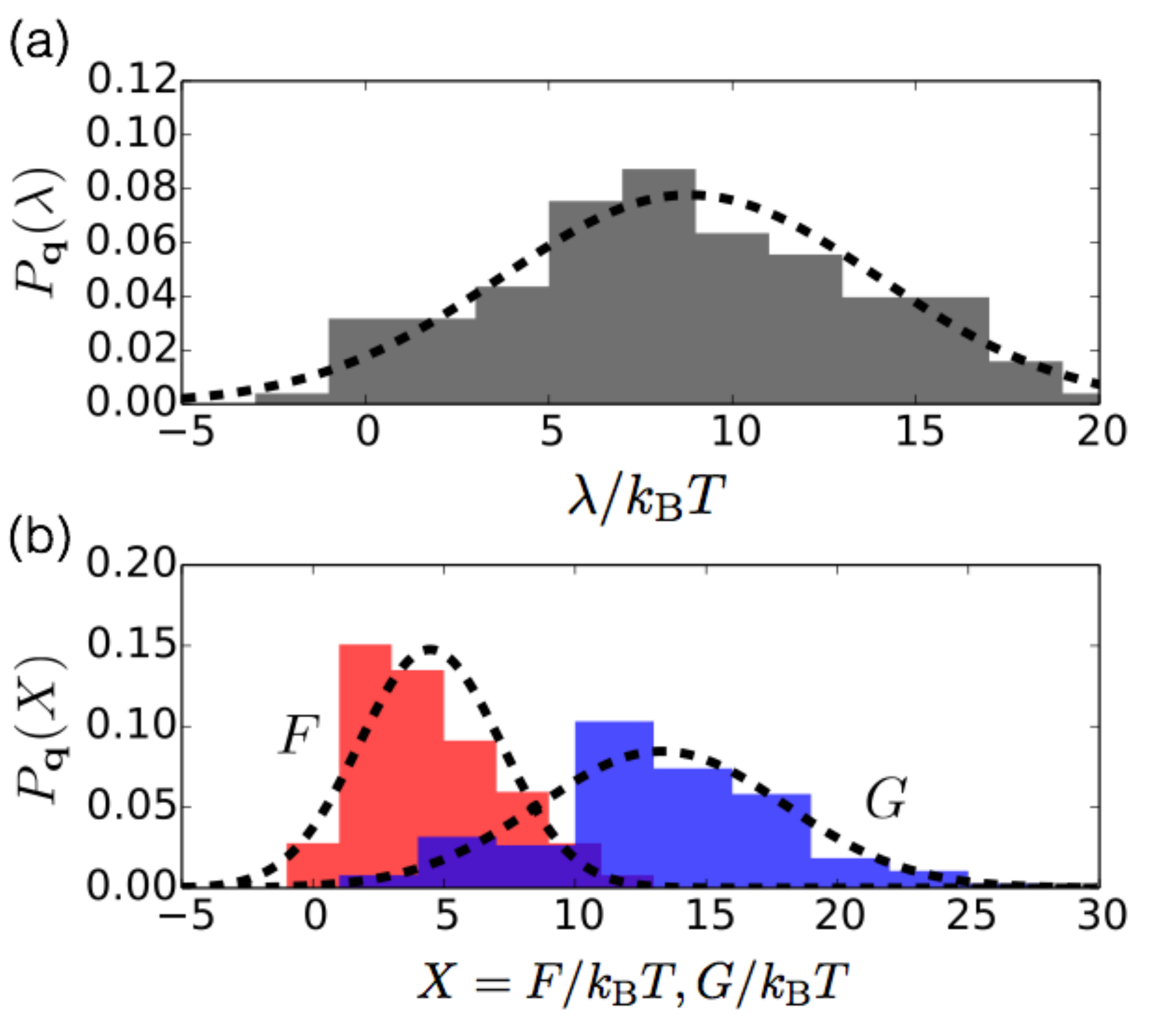}
\caption{(a) Probability distribution of surface reorganization energies, $\lambda$, for the 111 surface. The dashed line is a Gaussian fit. (b) Probability distributions of $F$, in red, and $G$, in blue for the 111 surface. Dashed lines are Gaussian fits with same means and variances.}
\label{Fi:9}
\end{center} 
\end{figure}

\section{Platinum 111 surface}

The main text focuses on the kinetics of the 100 surface, which is four fold-coordinated and more commensurate with extended hydrogen boning patterns resulting in unit monolayer coverage. We have also carried out this analysis for the 111 platinum surface, which is six-fold coordinated, and less commensurate with extended hydrogen bonding patterns resulting in an average coverage, which is nearly 80\%. Previous work has detailed differences in the monolayer dynamics and solvation properties between these two surfaces. Here, we focus on only those differences that alter pathways for desorption. 

First, a number of quantitative differences result when desorption occurs at the 111 crystal facet compared to the 100 facet. As shown in Fig.~\ref{Fi:8}, the timescales are generically shorter with a rate 2 times larger and a mean instanton time 2.5 ps.  These results are consistent with previous work that found faster relaxation times of water molecules adsorbed on the 111 compared to the 100. This is due to the six-fold coordination of the lattice with is less commensurate with hydrogen bonding that introduces a larger degree of disorder on the 111 surface and facilities motion.

These smaller characteristic timescales are mirrored in the relevant work quantities elucidated in the main text. Specifically, Fig.~\ref{Fi:9}a shows the distributions of $\lambda$ whose mean is similar to the 100 surface, but whose width is 1.25 times larger leading to a high probability of lower values.  Distributions of $F$ and $G$ are plotted in Fig.~\ref{Fi:9}b. The mean values of each for the 111 surface are shifted relative to the 100 by about 2 $k_\mathrm{B} T$. Within the limited statistics, these distributions are Gaussian. 

For the 111 surface, we find that the number of vacancies adjacent to a tagged site, increases monotonically with increasing reorganization energy. This is a reflection of the stable hexagonal ring structure that water forms on the surface, which is commensurate with the 6-fold coordination of the exposed lattice. For higher local surface densities, where tagged molecules do not sit adjacent to vacant sites, hydrogen bonding is frustrated and consequently these sites have a lower $\lambda$.

%

\end{document}